\documentclass[useAMS,usenatbib]{mn2e}
\usepackage{epsfig}
	
\title{Can large-scale structure probe CMB-constrained non-Gaussianity?}
\author[X. Kang, P. Norberg and J. Silk]
	{X. Kang$^1$\thanks{Email: kangx@astro.ox.ac.uk},
	  P. Norberg$^2$ and Joseph Silk$^1$ \\ 
        $^1$Department of Astrophysics, University of Oxford, Keble
	  Road, OX1 3RH, UK\\ 
        $^2$SUPA\thanks{The Scottish Universities Physics Alliance},
	  Institute for Astronomy, University of Edinburgh, Royal
	  Observatory, Blackford Hill, Edinburgh, EH9 3HJ,UK} 

\date{Draft version \today}

\def \lstar {L$^{*}_{\rm b_{\rm J}}$}
\def \hmpc {h$^{-1}$ Mpc}
\def \hmpccube {h$^{-3}$ Mpc$^{3}$}
\def \kms {s$^{-1}$ km}
\def \etal {et~al.}
\def \lcdm {$\Lambda$CDM}
\def\simlt{\lower.5ex\hbox{$\; \buildrel < \over \sim \;$}}
\def\simgt{\lower.5ex\hbox{$\; \buildrel > \over \sim \;$}}
\def\simpropto{\lower.2ex\hbox{$\; \buildrel \propto \over \sim \;$}}

\begin{document}

\maketitle

\label{firstpage}
 
%%%%%%%%%%%%%%%%%%%%%%%%%%%%%%%%%%%%%%%%%%%%%%%%%%%%%%%%%%%%%%%%%%%%%%%%%%%%%%%%%%%%%%%%%%%%%

\begin{abstract}
The first year Wilkinson Microwave Anisotropy Probe (WMAP) set
quantitative constraints on the amplitude of any primordial
non-Gaussianity. We run a series of dark matter-only N-body
simulations with the WMAP constraints to investigate the effect of the
presence of primordial non-Gaussianity on large scale structures. The
model parameters can be constrained using the observations of
protoclusters associated with Ly-$\alpha$ emitters at high redshift
($2 \leq z \leq 4$), assuming the galaxy velocity bias can be modelled
properly.  High redshift structure formation potentially provides a
more powerful test of possible primordial non-Gaussianity than does
the CMB, albeit on smaller scales.  Another constraint is given by the
local galaxy density probability distribution function (PDF), as
mapped by the 2 degree Field Galaxy Redshift Survey (2dFGRS). The PDF
of 2dFGRS \lstar\ galaxies is substantially higher than the standard
model predictions and requires either a non-negligible bias between
galaxy and dark matter on $\sim 12$~\hmpc\ scales or a stronger
non-Gaussianity than allowed by the WMAP year one data. The latter
interpretation is preferred since second-order bias corrections are
negative. With a lower normalisation of the power spectrum
fluctuations, $\sigma_8=0.74$, as favoured by the WMAP 3 year data,
the discrepancy between the Gaussian model and the data is even
larger.

\end{abstract}

\begin{keywords}
cosmology: theory -- large-scale structure of the Universe --
galaxies: clusters: general 
\end{keywords}

%%%%%%%%%%%%%%%%%%%%%%%%%%%%%%%%%%%%%%%%%%%%%%%%%%%%%%%%%%%%%%%%%%%%%%%%%%%%%%%%%%%%%%%%%%%%%

%%%%%%%%%%%%%%%%%%%%%%%%%%%%%%%%%%%%%%%%%%%%%%%%%%%%%%%%%%%%%%%%%%%%%%%%%%%%%%%%%%%%%%%%%%%%%

\section[]{Introduction}
\label{sec:Intro}
 
The best limits on possible initial non-Gaussianities in the
primordial density fluctuation distribution have hitherto been imposed
by the CMB (Komatsu \& Spergel 2001).  There are many models which can
generate such non-Gaussianities, including multiple field inflation or
more speculative models such as those assuming non-homogeneous
reheating processes or those seeded with topological defects.
Moreover such non-Gaussianities may be scale-dependent, and it is
therefore important to complement CMB searches with smaller scale
probes, on cluster and galaxy scales.

Our approach here is phenomenological, following that of Mathis, Diego
\& Silk (2004), and focuses on the large-scale galaxy distribution.
Tracers of primordial non-Gaussianities include possible large-scale
power excesses in the nearby galaxy distribution as mapped by 2dFGRS
and SDSS redshift surveys, the early formation of protoclusters up to
$z\sim4$ as traced by radio galaxies surrounded by Ly-$\alpha$
emitters and Lyman-break galaxies, and the possible excess of angular
power in CMB temperature fluctuations seen by CBI and ACBAR on cluster
scales (which is most likely due to the thermal SZ effect). With the
claimed reduction in normalisation of the fluctuation power spectrum
to $\sigma_8=0.74\pm 0.05$ (Spergel \etal\ 2006), such questions take
on renewed urgency.

Our approach here is to study the implications via dark matter N-body
simulations of a simple non-Gaussian model, which modifies the
frequency of rare peaks in the primordial density field that defines
the locations where massive clusters (or superstructures) are
formed. If the probability distribution function of the density field
shows an excess (or deficit) of high peaks compared to the Gaussian
case, protoclusters of given mass form earlier (or later), and systems
above a given mass will be more (or less) abundant at fixed
redshift. Sadeh, Rephaeli \& Silk (2006) presented analytical models,
whereas here we simulate and evolve a suite of N-body simulations of
300~\hmpc\ on the side.

%%%%%%%%%%%%%%%%%%%%%%%%%%%%%%%%%%%%%%%%%%%%%%%%%%%%%%%%%%%%%%%%%%%%%%%%%%%%%%%%%%%%%%%%%%%%%

\section[]{The case for non-Gaussianity}
\label{sec:Simus}

We consider a simple non-Gaussian model, where the amplitude
of non-Gaussianity in the primordial fluctuations is parametrised by a
non-linear coupling parameter, $f_{NL}$ (Komatsu \etal\ 2003). We refer
to this non-Gaussian model as the $f_{NL}$ model. The Bardeen curvature
perturbations $\Phi$ at any given position $x$ can be written as  
\begin{equation}
\Phi(x) = \Phi_{L}(x) + f_{NL}[\Phi_{L}^{2}(x)-<\Phi_{L}^{2}(x)>]
\label{eq:phi}
\end{equation}
where $\Phi_{L}(x)$ are Gaussian linear perturbations with zero
mean. This parameterisation is useful for setting  quantitative
constraints on the amount of non-Gaussianity allowed by the CMB
data. Komatsu \etal\ (2003) used first year WMAP data (WMAP-1) to
establish limits on $f_{NL}$ and obtained $-58<f_{NL}<134$ at 95\%
confidence.

We produce the initial conditions for the $f_{NL}$ simulations using
the conventional grid method, but when calculating the displacement
using the Zel'dovich approximation, we use the perturbed newtonian 
potential, which is $-1$ times the curvature perturbation
given by Equation~\ref{eq:phi}. We then run a series of dark matter
N-body simulations using the publicly available code GADGET (Springel,
Yoshida \& White 2001). We consider three values for $f_{NL}$
(-58,0,134) consistent with WMAP1 constraints, where $f_{NL}=0$
corresponds to the standard Gaussian model. For these simulations, we
impose the initial power spectrum to be given by an adiabatic CDM
transfer function with spectral slope $n_{s}=1$. We use the `old`
concordance cosmological model, {\it i.e.} $\Omega_{m}=0.3,
\Omega_{\Lambda}=0.7$, $h=0.7$ and $\sigma_{8}=0.9$. Recent results
from WMAP-3 prefer a lower value for $\sigma_{8}$ (Spergel \etal\
2006). We therefore run another series of simulations using the WMAP-3
cosmological parameters (ie. $\Omega_{m}=0.238, \Omega_{\Lambda}
=0.762$, $h=0.73$, $\sigma_{8}=0.74$), both for $f_{NL}=0$ and
$f_{NL}=134$. All N-body simulations are started at redshift
$z_{init}=70$ and we follow $128^3$ particles in a cube of 300~\hmpc\
aside, implying a particle mass of $\rm \sim 10^{12}h^{-1}M_{\odot}$,
which, for the present study, gives high enough resolution as shown in
Section~\ref{sec:Res}. For each of the above models, we present the
mean results from ten statistically equivalent
realizations. Hereinafter, we conservatively discuss the results for
the cosmological model with $\sigma_{8}=0.9$ unless otherwise stated.

\section{Results}
\label{sec:Res}
\subsection{Halo mass function in non-Gaussian models}

\begin{figure}
\begin{minipage}{8cm}
\epsfig{file=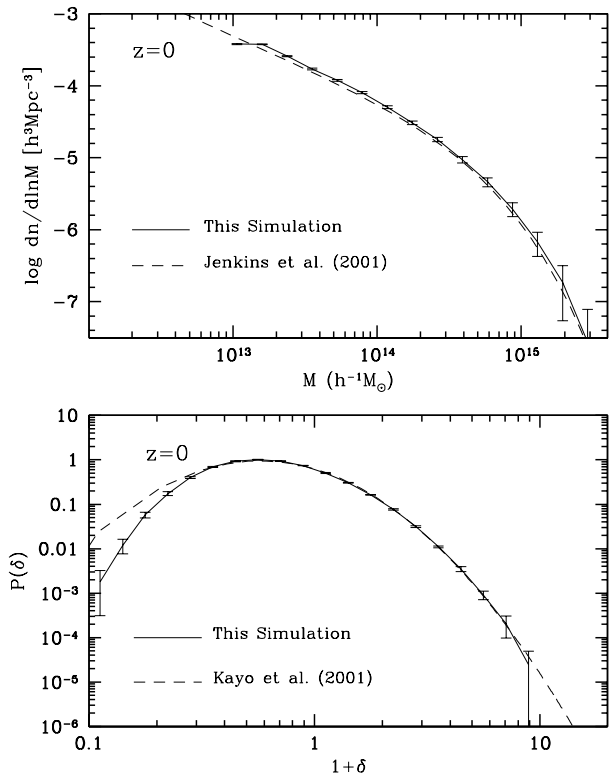,width=8cm}
\caption{Upper panel: mass function of dark matter halos at z=0 from 
simulations with Gaussian initial conditions. The solid line shows the mean
mass function from 10 simulations with Gaussian initial conditions,  
and the dashed line shows the Jenkins et al. (2001) halo mass function. 
Lower panel: the probability distribution function of the density contrast. 
Again, the solid line shows the mean from our 10 Gaussian simulations and the
dashed line is the corresponding analytic prediction of Kayo et al. (2001).}
\label{fig:compare}
\end{minipage}
\end{figure}

As with any simulation run, it is essential to check if our analysis
is affected by the chosen mass resolution. In Fig.~\ref{fig:compare},
we show a comparison between our simulations and results from either
other high-resolution N-body simulations or analytical models. The
upper panel compares the halo mass function of our Gaussian
simulations with the fitting formula of Jenkins et al. (2001), which
is based on a series of high-resolution N-body simulations with
similar cosmology.  Our estimate of the halo mass function is in good
enough agreement with the Jenkins et al. (2001) fit.  In the lower
panel, we show the probability distribution function (PDF) of the
density contrast smoothed with a spherical top-hat filter of radius
12~\hmpc.  Kayo et al. (2001) have shown that the PDF can be well
described by a lognormal distribution. Here we show the mean PDF from
our 10 Gaussian simulations as the solid line, and the errors
correspond to the estimated variance. The dashed line shows the
analytic relation given by Kayo et al. (2001), which is shown to be
accurate at $\delta \geq -0.5$.  It can be seen that our simulation
matches well the analytic prediction.  The main point of the plot is
to show that our mass resolution is good enough to properly resolve
the over-dense tail of the probability distribution function, for a
typical smoothing scale used later on in our analysis.

\begin{figure}
\begin{minipage}{8cm}
\epsfig{file=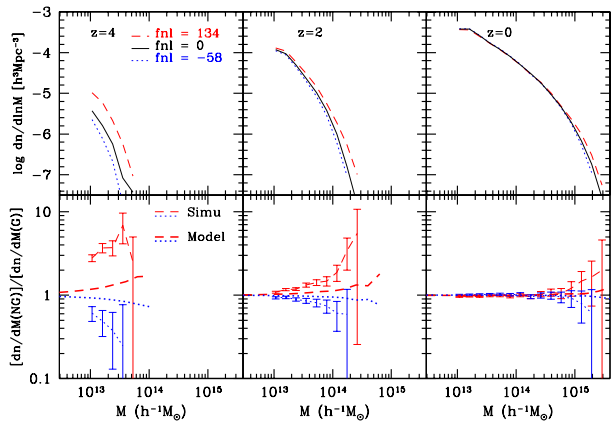,width=8cm}
\caption{Upper panel: dark matter halo mass functions with $f_{NL}$
  parameters 134 (red dashed), 0 (black solid), -58 (blue
  dotted). Lower panel: the ratio of halo mass functions of
  non-Gaussian and Gaussian models. Thick lines with error bars are
  for our suite of DM simulations and thin lines are the analytic
  model predictions of Matarrese et al. (2000), for the same set of
  $f_{NL}$ parameters (same line styles and colours as in the upper
  panel).}
\label{fig:MF}
\end{minipage}
\end{figure}

Our main motivation is to see whether the type of non-Gaussian models
considered here, ie. f$_{NL}$ models, can produce more massive
structures than does the standard Gaussian model. In Fig.~\ref{fig:MF}
we present the evolution of the dark matter halo mass function as a
function of redshift. The upper panel shows the results with WMAP-1
constrained f$_{NL}$ model parameters. A distinctive feature is that a
positive f$_{NL}$, which gives a more strongly disturbed gravitational
field, results in the formation of increased numbers of virialized
structures. This effect is also much stronger at higher redshifts, as
expected from the theoretical models of Matarrese et al. (2000). In
the lower panel of Fig.~\ref{fig:MF}, we show the ratio of halo mass
functions of non-Gaussian, $dn/dM(NG)$, and Gaussian models, $dn/dM(G)$,
by the thick lines with error bars from our ten statistically
equivalent realizations.  Even though the simulation predicts the ratio
of halo mass function to increase with halo mass and with increasing
redshift, there is still a large discrepancy with the analytical model
expectations of Matarrese et al. (2001).  Our simulations produce a
much larger number density of dark matter halos for non-Gaussian
models than expected by Press-Schechter-based model predictions.  The
main difference between the Press-Schechter halo mass functions and
the N-body mass functions can be attributed to a different treatment
of the dynamics of halo formation (cf. Komatsu et al. 2003).  It is
valid to ask whether primordial non-Gaussianity also changes the
dynamics of halo formation. This issue is unfortunately beyond our
current study, as following the dynamics of halo formation requires
much higher-resolution simulations than the ones we have run, together
with more regularly spaced simulation outputs. We therefore delay this
aspect of non-Gaussianity to future work.

\subsection{Protoclusters at z $\sim$ 2-4}

As previously shown, there are significant differences between the
Gaussian model and the $f_{NL}$ models at high redshift, in particular
for large positive $f_{NL}$ values, which results in more structure
being formed. Miley \etal\ (2004) have reported a protocluster at
$z=4.1$ around a Ly-$\alpha$ emitter galaxy. They found 21
spectroscopically confirmed galaxies around the Ly-$\alpha$ emitter
galaxy, and the 1-dimensional velocity dispersion among these galaxies
is about 325~\kms.  Kurk \etal\ (2004) also observed a protocluster at
$z=2.16$ with a velocity dispersion of $\sim 360$ s$^{-1}$ km over 3 to 5
Mpc, corresponding to co-moving sphere radii of 5 to
9~\hmpc. Following Robinson \& Baker (2000) and Mathis, Diego \& Silk
(2004), we sample such structures in our simulations by throwing down
a large number of spheres with radii of 7~\hmpc\ and calculate their
1-D dark matter velocity dispersion.

\begin{figure}
\begin{minipage}{8cm}
\epsfig{file=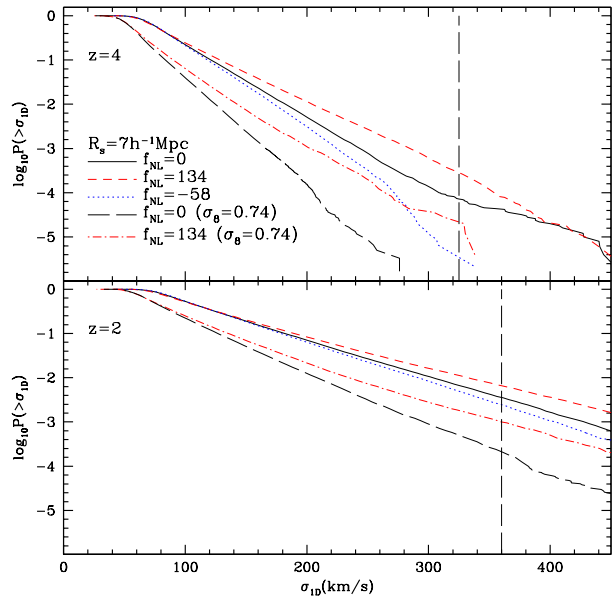,width=8cm}
\caption{Cumulative probability distribution function of the
  one-dimensional velocity dispersion measured within spheres of radii
  7~\hmpc. Upper (lower) panel show the $z=4$ ($z=2$) results. Vertical
  dashed lines indicate the observed velocity dispersion by Miley \etal\
  (2004) and Kurk \etal\ (2004).}
\label{fig:PDV}
\end{minipage}
\end{figure}

In Fig.~\ref{fig:PDV} we show the cumulative probability distribution
function of the measured 1-D dark matter velocity dispersion. In the
top panel, we compare our findings with the measurement of Miley
\etal\ (2004) at $z\sim4$ .  In the bottom panel, we plot the
equivalent curves at $z=2$ and compare with the velocity dispersion
measurement of Kurk \etal\ (2004).  Contrary to Mathis \etal\ (2004),
we find that our Gaussian simulation with $\sigma_{8}=0.9$ can explain
the existence of structures with velocity dispersions larger than
320~\kms, such as the protocluster of Miley \etal\ (2004). Although
the simulation of Mathis \etal\ (2004) used the same cosmological
parameters as our simulation, they adopt too small a simulation volume
(box of 100~\hmpc\ aside), with the consequence that the cut-off in
the input power spectrum on large scales depresses the velocity
dispersion in their simulation.

Fig.~\ref{fig:PDV} shows that for all simulations with $\sigma_8=0.9$,
it is possible to {\it observe} a $z\sim4$ protocluster with 1-D
velocity dispersion of about 320~\kms. On the other hand in a low
$\sigma_{8}$ universe (like one with $\sigma_{8}=0.74$, as favoured by
WMAP-3), the chance of {\it observing} such protoclusters is very
unlikely, unless there is a strongly disturbed primordial non-Gaussian
field (such as one with $f_{NL}=134$). We note that these are not in
any case very strong constraints, as we have ignored any velocity bias
between the galaxies and the dark matter, which to some extent could
solve the apparent discrepancy. Therefore we have not attempted to
quantify the predicted number density of such objects at $z\sim4$. Only
detection of a large number of protoclusters of similar size or bigger
at $z\sim4$ could actually put real constraints on this simple model,
together with some theoretical constraints on the strength of velocity
bias, as would be measurable in higher resolution N-body simulations.
The lower panel of Fig.~\ref{fig:PDV} shows that, at $z\sim2$, it is
rather easy to find, in all of the simulations considered, regions
with velocity dispersion larger than 400~\kms. This highlights the
fact that only very high redshift protoclusters are likely to set
constraints on the cosmological parameters using the 1-D velocity
dispersion.

\begin{figure}
\begin{minipage}{8cm}
\epsfig{file=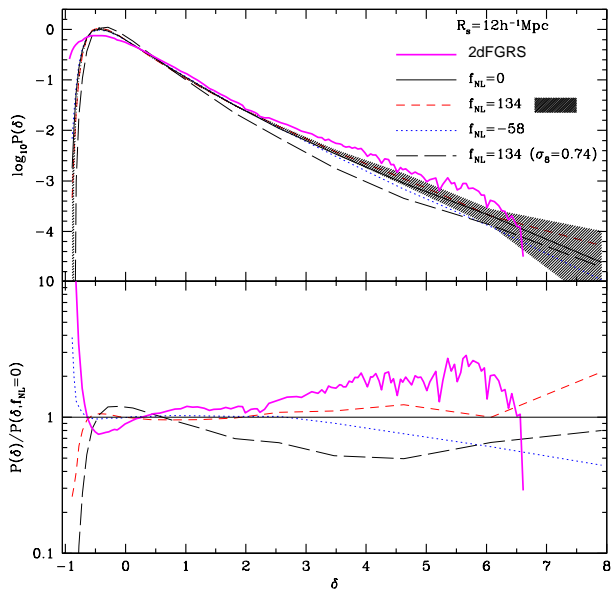,width=8cm}
\caption{Top panel: Probability distribution function of the
  mass (galaxy) density contrast within spheres of 12~\hmpc\ radii:
  four sets of dark matter PDFs from Gaussian and non-Gaussian models
  (see labels in panel) and, in magenta, the PDF of 2dFGRS \lstar\
  galaxies. Lower panel: the ratio of the PDFs shown in the top panel
  to the Gaussian PDF. This explicitly shows that the PDF of
  2dFGRS \lstar\ galaxies is distinctively different to any of the DM
  models presented and highlights particularly well that a lower
  $\sigma_8$ value increases the discrepancy between data and models
  even further.}
\label{fig:PDF_z0}
\end{minipage}
\end{figure}

\subsection{Large scale structures in the 2dFGRS}
\label{sec:lss}

A clear limitation of the preceding results is the sparseness and
quality of the data that we are comparing to. Hence we consider in
this section constraints from local large-scale structure results.
We estimate the galaxy probability distribution
function (PDF) using a 2dFGRS \lstar\ volume-limited
sample (Norberg \etal\ 2002; Baugh \etal\ 2004), comparable in volume
to a simulation cube of $\sim$~200~\hmpc\ on the side. The PDF is 
measured by  Fast-Fourier-Transform (FFT) smoothing of the galaxy 
distribution with a spherical top-hat filter of radius $R.$  For the 
smoothing scales considered (4 to 16~\hmpc), we recover the results 
of Croton \etal\ (2004), who used a count-in-cells (CiC) method to 
measure the galaxy PDF. The FFT-based method, once we accurately 
account for edge effects (particularly important when 
smoothing over large scales), is several orders of magnitude 
faster than standard CiC.

In the top panel of Fig.~\ref{fig:PDF_z0}, we show the 2dFGRS \lstar\
galaxy PDF, smoothed over 12~\hmpc, together with the dark matter PDF
of the $f_{NL}$models specified in the caption.  In the bottom panel,
we show the ratio between the PDFs presented in the top panel and the
PDF of the Gaussian simulation.  Clearly the 2dFGRS galaxy PDF lies
well above all model predictions, and in particular it is very
discrepant with models using a low $\sigma_8$ value, as currently
favoured by WMAP-3 data.

\begin{figure}
\begin{minipage}{8cm}
\epsfig{file=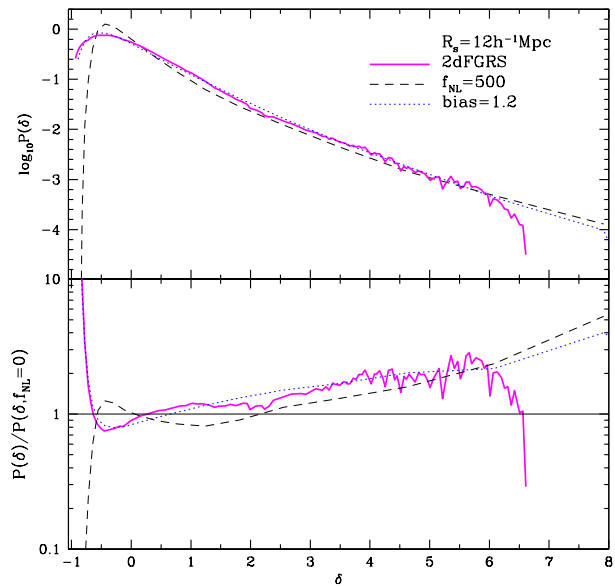,width=8cm}
\caption{Top panel: Probability distribution function of the
  mass (galaxy) density contrast within spheres of 12~\hmpc\ radii: in
  black, the PDF of a $f_{NL}=500$ model; in blue, a biased \lcdm\
  PDF; in magenta, the PDF of 2dFGRS \lstar\ galaxies. Both models
  assume $\sigma_8=0.90$ cosmology. Lower panel: the ratio of the PDFs
  shown in the top panel to the Gaussian \lcdm\ PDF. For a high
  $\sigma_8$ value, both models seem able  to reproduce the tail of
  the data PDF rather well.}
\label{fig:PDF_galaxy}
\end{minipage}
\end{figure}

Two possible approaches may help reconcile the models with the
observations. Firstly, in Fig.~\ref{fig:PDF_z0}, we compare the galaxy
PDF with the mass PDF, ignoring any bias between them. We consider a
simple linear bias model (b $\simeq 1.2$) between galaxy and dark
matter on the weakly non-linear scales of 12~\hmpc\ in the Gaussian
model.  Secondly, we consider a more extreme positive non-Gaussian
model with $f_{NL}=500$, which predicts a significantly higher
probability in the tail of the PDF. The constraints from WMAP on
$f_{NL}$ are only for very large scales, and the non-Gaussianity may
be scale-dependent. In Fig.~\ref{fig:PDF_galaxy}, we compare these two
alternative models with the 2dFGRS \lstar\ PDF.

In the following discussion, we assume that the data extracted from
the 2dFGRS is a fair representation of the Universe, {\it i.e.} we
require our model to easily reproduce the observations, something one
potentially expects from a volume close to $8\times
10^{6}$~\hmpccube. Clearly, if $\sigma_8 = 0.9$, then both a standard
$\Lambda$CDM with relatively strong linear bias on non-linear scales
($b \simeq 1.2$) or a strong non-Gaussian model ($f_{NL}=500$) without
bias reproduce the 2dFGRS PDF rather well.  In the case of a low
$\sigma_8$ Universe, the bias needed for a pure Gaussian model would
be uncomfortably large ($b \simeq 1.46$), especially when accounting
for the most recent determination of galaxy bias on weakly non-linear
scales using the same 2dFGRS samples (Gazta$\tilde{\rm n}$aga \etal\
2005). Alternatively, a strong positive non-Gaussianity ($f_{NL}\sim
500$) coupled with a moderate galaxy bias could resolve the
discrepancy between data and models.  In fact, a linear bias model on
the scales of 12~\hmpc\ may not be realistic. Gazta$\tilde{\rm n}$aga
et al. (2005) have recently measured the galaxy bias on weakly
non-linear scales (ie. $\sim$~10 to $\sim$~30~\hmpc) using the 2dFGRS
samples and found that the bias is non-linear in over-dense regions
($\delta \gg 1$).  However the second order corrections are found to
be negative, lowering the galaxy density, and hence favouring the
non-gaussian interpretation.

Two issues are worth mentioning in this analysis: we focus in
particular on the \lstar\ volume-limited sample, because of the
presence of two very large coherent superstructures in each 2dFGRS
regions
\footnote{The Northern one is part of the SDSS great wall}, which
strongly influences all higher-order statistics (Baugh \etal\ 2004;
Croton \etal\ 2004; Gazta$\tilde{\rm n}$aga \etal\ 2005; Nichol \etal\
2006), and even to some extent the 2-point correlation function. The
other reason is that \lstar\ galaxies are expected to  only be very
mildly biased tracers of the underlying dark matter (eg. Verde \etal\
2002; Gazta$\tilde{\rm n}$aga \etal\ 2005). Results from 
a larger volume-limited sample such as the $\sim2.5$~\lstar\
sample are more difficult to interpret, mostly because the galaxy
density is  much lower, making the tail of the PDFs harder to
measure. 

\section[]{Conclusions}
\label{sec:CCL}

In this paper, we have studied a simple non-Gaussian model in which
the initial potential has a small perturbation relative to the
Gaussian random field and can be described by a simple non-linear
coupling parameter (Komatsu \etal\ 2003). We produce a series of
N-body simulations to model structure formation with both Gaussian and
non-Gaussian initial conditions. We estimate the predicted halo mass
function as a function of redshift, and make comparison with
observations for the PDF of the 1-D DM velocity dispersion as well as
for the PDF of the DM overdensity.

We find that, compared to the Gaussian model, a non-Gaussian model
with deeper potential wells (ie. positive $f_{NL}$) produces larger
structures, whereas models with negative $f_{NL}$ values do the
opposite.  The differences are very pronounced at high redshift, but
are less evident at $z \sim 0$. The presence of large coherent
structures at high redshift is a better indicator of early structure
formation than similar objects at low redshift. Under the assumption
of no velocity bias between galaxy and mass, our results show that it
would be difficult to observe the $z\sim4.1$ protocluster of Miley
\etal\ (2004) in a $\sigma_{8}=0.74$ Universe with Gaussian initial
conditions, but becomes much easier with a strong non-Gaussian field
(eg. $f_{NL}=134$).  Protoclusters at high redshift can put strong
constraints on non-Gaussian models, once the effects of bias and
sample selections are properly addressed.  High redshift structure
formation potentially provides a more powerful test of possible
primordial non-Gaussianity than does the CMB, albeit on smaller
scales.

This of course assumes that we live in a typical region of the
universe.  However the large-scale coherent structures observed in the
2DFGRS (Baugh et al. 2005) that dominate the higher order statistics,
and also seen in the SDSS, might indicate that a larger volume of the
universe must be sampled to robustly test non-Gaussianity.

A detailed comparison between the 2dFGRS \lstar\ galaxy PDF and the
PDFs from a series of dark matter models show how difficult it is to
reproduce the observed high tail in the galaxy PDF, when smoothed on
12~\hmpc\ scales. In a $\sigma_8=0.90$ Universe, we argued that only a
Gaussian \lcdm\ model with strong linear bias between galaxies and
mass ($b \sim 1.2$) or a very strongly non-Gaussian model, with
$f_{NL} \sim 500$, can reconcile the results mapped by the 2dFGRS. The
bias constraints obtained by Gazta$\tilde{\rm n}$aga \etal\ (2005)
using the same 2dFGRS samples favour the latter option. However the
complication of galaxy bias on weakly non-linear scales in over-dense
regions ($\delta \gg 1$) makes it difficult to constrain the model
parameters based on a simple linear bias model. Nevertheless, in a low
$\sigma_8$ Universe as favoured by recent WMAP-3 results, the case for
primordial non-Gaussianity is strengthened.  Further tests, with
realistic galaxy formation recipes, are needed to properly address
these issues.  Our conclusions complement those in a recent
theoretical paper by Hikage, Komatsu \& Matsubara (2006), namely that
non-Gaussianity can be well constrained by high redshift galaxy
surveys and that non-linear galaxy bias makes it difficult to detect
primordial non-Gaussianity at low redshift.

\section{Acknowledgements}
The authors would like to thank the referee for useful comments.
XK acknowledges a Royal Society Chinese Fellowship for financial 
support. PN acknowledges receipt of a PPARC post-doctoral fellowship
held at the IfA. XK and PN are grateful for the hospitality of the
Physics Department of the University of Oxford during respective
stays and visits.  

\bsp \bibliographystyle{mnras}

\end{document}